\documentclass[preprint]{aastex}

\def\eg{{\rm e.g.~}}  
\def\etal{{\rm et~al.~}} 
\def\lapp{\ifmmode\stackrel{<}{_{\sim}}\else$\stackrel{<}{_{\sim}}$\fi}
\def\gapp{\ifmmode\stackrel{>}{_{\sim}}\else$\stackrel{>}{_{\sim}}$\fi}        

\begin{document}

\title{Understanding the Fundamental Plane and the Tully Fisher Relation} 

\author{Jeremy Mould$^1$}
$^{1}$ARC Centre for Dark Matter Particle Physics and Centre for Astrophysics and Supercomputing, Swinburne University, Hawthorn, Vic 3122, Australia



\begin{abstract}
The relation between early type galaxy size, surface brightness and velocity dispersion, ``the fundamental plane",
has long been understood as resulting from equilibrium in their largely pressure supported stellar dynamics.
The dissipation and feedback involved in reaching such an equilibrium through merger formation of these galaxies over cosmic time
can be responsible for the orientation of the plane. 
We see a correlation between surface brightness enhancement and youth in the 6dF  Galaxy Survey.
Correlations of this `tilt' with stellar mass, age, concentration, shape and metallicity now point the direction for further work
on the resolved kinematics and structure of these nearby galaxies and on their initial mass function and dark matter component.

On the face of it, the Tully Fisher relation is a simpler one dimensional scaling relation. However, as late type galaxies have bulges as well as disks,
and, as the surface density of disks is only standard for the more massive
galaxies, additional parameters are involved. 

\end{abstract}

\section{Introduction}
The fundamental plane (FP) of early type galaxies is fairly aptly named, because in the 3-space of velocity dispersion, ($\sigma$), surface brightness and effective radius is expressed the dynamical equilibrium that a galaxy has reached, but not (the interesting part) how it got there. It has long been understood that the virial theorem\footnote{An anonymous referee has pointed out that the
Virial Theorem does not hold at any arbitrary radius in a system in dynamical equilibrium. It holds for a particular radius. This is shown in work on mass estimators, for example by Wolf et al (2010) and Walker et al (2009).} will place quiescent galaxies with a well defined mass to light ratio in a plane in this space. The observed tilt of the plane differs from the expectation, however, and the deviation can be patched using the halo occupation distribution formalism (e.g. Moster et al 2010, Mould 2014, 2017). But that formalism is not physics: it is a mapping between the dark matter universe and the observed one. The reason for the copout, of course, is the difficulty of modelling baryonic processes such as star formation with their challenging demands for spatial resolution. 


\section{Building the FP}
If mergers are the manner in which early type galaxies are formed, a schematic
understanding can be based on the serial accretion by a galaxy with mass, kinetic and potential
energies (m, T, V)  of  dwarfs ($\delta m,~ \delta T,~ \delta V$).
Each is in virial equilibrium\footnote{Or in a tilted virial plane}, so that 2T = --V and 2$\delta T ~=~ -\delta V$.
If energy D is dissipated in the interaction, the energy equation is

$$\frac{_1}{^2} \delta m \sigma_0^2 + \frac{_1}{^2} m \sigma^2 + D = \frac{_1}{^2} (m + g \delta m) (\sigma + \delta \sigma)^2 \eqno(1)$$

\noindent where $\delta T ~=~ \frac{1}{2} \delta m \sigma_0^2$  and T = $\frac{1}{2} m \sigma^2$
and g is unity if mass is conserved.
\\
\\
If $D$ = $fGm\delta m/r_e$, where $f$ is a free parameter and $r_e$ is the effective radius, then

$$\sigma \delta \sigma (m + g \delta m) = \frac{_1}{^2} \delta m (\sigma_0^2 - g\sigma^2) +   fGm\delta m/r_e\eqno(2)$$

\noindent which yields the differential equation

$$\frac{d \log \sigma} {d \log m} = \frac{_1}{^2} (\frac{\sigma_0^2} {\sigma^2} - g) + \frac{r_0}{r_e} \eqno(3)$$

\noindent where $r_0$ has replaced  $f$ to keep a dimensionless equation simple.\\
\\

We integrate equation (3) numerically and adopt M/L~$\propto~\sigma^\epsilon$, following Mould (2014), and $r_e ~=~ r_0 \sigma^{1.75}$ plus a dispersion,
to obtain the fundamental plane shown in Figure~ 1. 
In our adopted M/L $\epsilon$ is not a free parameter\footnote{Cappellari et al (2013a) find $\epsilon$ = 0.72.} The proportionality constant will contain stellar population dependences, such as metallicity.
The fit to Figure 1 is $r_e ~\propto~ \sigma^{1.6} SB^{-0.4}$, where SB is surface brightness, which is not too dissimilar to that
found for 6dF by Magoulas \etal (2012): $r_e ~\propto~ \sigma^{1.53} SB^{-0.89}$.
\begin{figure}
\includegraphics[scale=0.5]{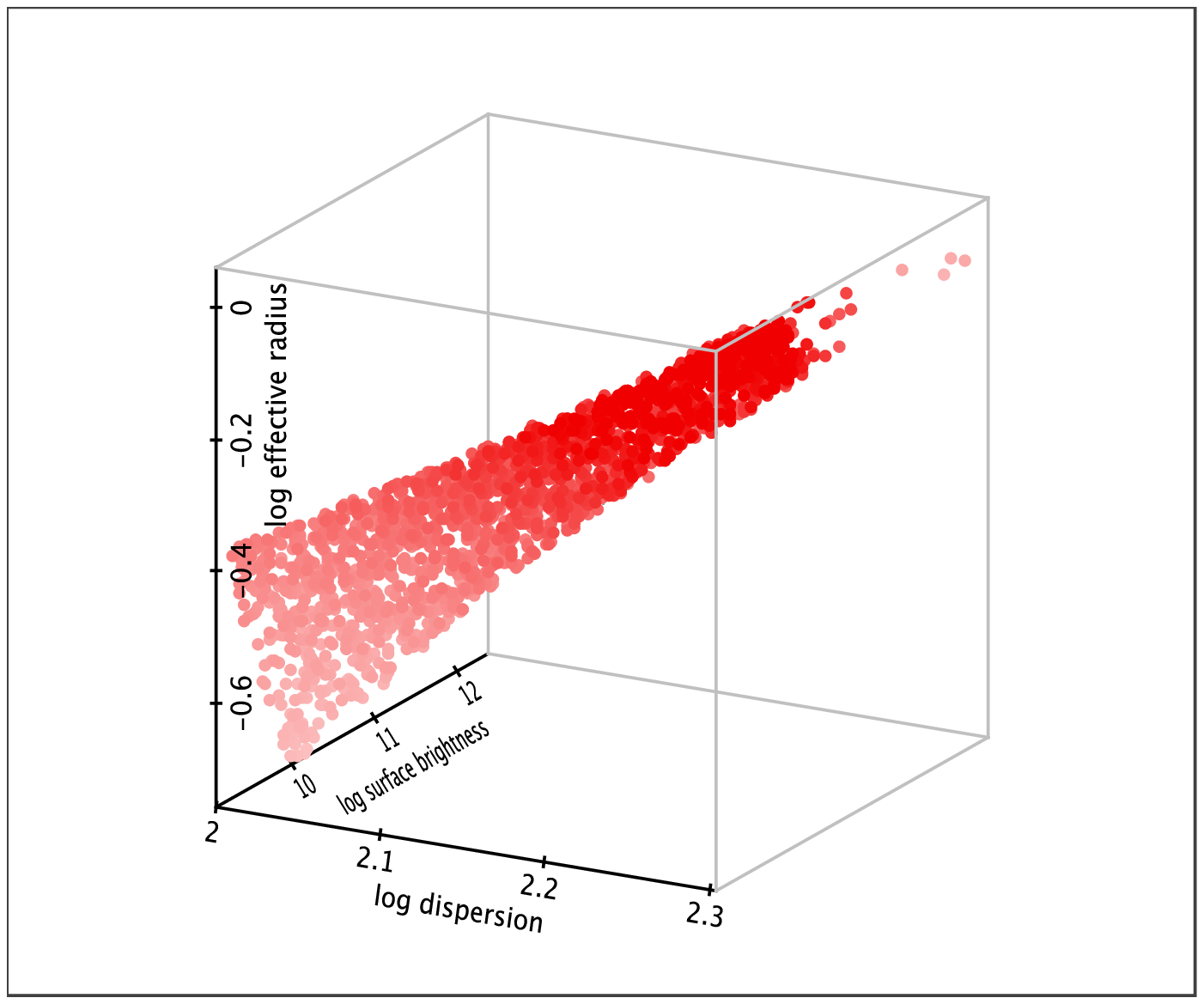}
\caption{A simulated FP obtained by integrating equation (3).}
\end{figure}

The foregoing is just a schematic insight into the FP, illustrating how a tilt from 
the virial plane may arise from dissipation, parameterized by f and feedback/mass loss parameterized by g,
both of which could be functions of scale\footnote{A particular choice of f($\sigma$) \& g, for example, yields the Faber Jackson (1976) power law relation.}. 
With two free parameters it is not surprising
that the observed FP can be approximated. But the point is, they stand for two processes which
must be fully physically modelled if the FP is to be understood.

\subsection{Bulges and nuclei}
The 6dFGS (Jones et al 2009) selected early type galaxies from 2MASS (Skrutskie \etal 2006).
The sample therefore contains ellipticals, S0s and spiral galaxies with large bulges. Bulges are believed to form in a hybrid manner, 
some more like ellipticals, others as a result of the secular evolution of disks (Kormendy \& Kennicutt 2004).
In either case the formalism of this section applies and it is not surprising that all of these galaxies fall on a common FP. 
Pseudobulges, as secular evolution dominated bulges are called, distinguish themselves by the
low bulge fraction of the total light, flatter shapes, high ratio of rotation
to velocity dispersion, bars, small Sersic index and tendency towards exponential
surface brightness profiles. We examine shape (or ellipticity) and
Sersic index, n, in $\S$4.

We note that, within bulges, nuclei form by a similar accretion process and are said
to co-evolve with bulges (\eg Kormendy \& Ho 2013). So it is not surprising
to see a power law relation between nuclear black hole mass and velocity 
dispersion (Magorrian \etal 1998, Ferrarese \& Merritt 2000, Graham 2012), again following the formalism of the preceding section,
leading to a power law relation. 
Spectra of active galactic nuclei arising from 6dFGS were discussed by Masci \etal (2010).

\section{Stellar mass}
The 6dFGS sample (Campbell \etal 2014) at z $\approx$ 0 has the merit of considerable
volume and southern complementarity to SDSS.

To describe the deviation from the virial plane, following D'Onofrio \etal (2013), we take

$$\Delta_{FP} = (a-2) \log \sigma + (b-1) \log SB \eqno(4)$$

\noindent where $a$ and $b$ are FP coefficients 
given in that reference.

$$r_e \propto \sigma^a SB^b \eqno(5)$$

\noindent Springob \etal (2012) and Proctor \etal (2008) measured Lick indices for the galaxies, giving estimates
of metallicity, $Z$, and age, $t$, for their stellar populations. Here and in Magoulas et al (2012) 
we use those spectra for which the quality factor is 10 or less and S/N $>$ 9.
Maraston (2005) models\footnote{http://www.icg.port.ac.uk/\\$\sim$maraston/Claudia\%27s\_Stellar\_Population\_Model.html}
 for the 2MASS J band 
adopted in the Campbell \etal (2014) dataset predict $M_{model}/L ~=~ M/L(Z,t)$
for Salpeter IMF and an assumed horizontal branch distribution.

\begin{figure}
\includegraphics[scale=0.35, angle=-90]{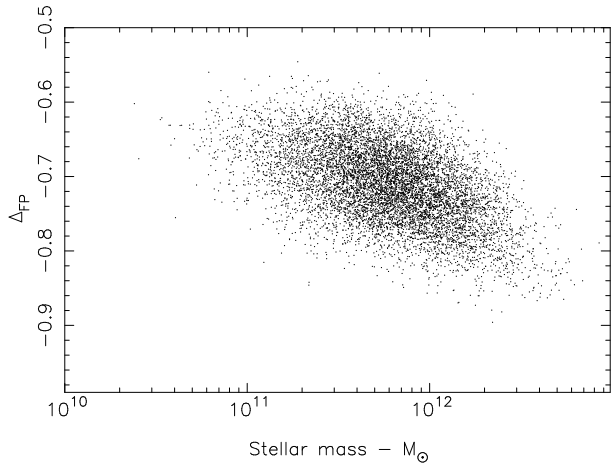}
\caption{Deviation from the virial plane. 8971 6dF galaxies.  
The values of $a$ \& $b$ defining the virial plane are those of
Magoulas et al (2012).
An IMF dependence
on stellar mass could be a key influence here (Conroy \etal 2012, 2013),
but the real situation could be more complex (see text).}
\end{figure}

Based on these mass-to-light ratios, Figure 2 is the deviation from the virial plane as a function of stellar mass.
The primary tilt of the FP is large and dependent on the mass of the galaxy.
Its origin is a hybrid of M/L dependence on mass and structural variation with mass.
In this context M/L refers to both the dark and baryonic matter components.




\subsection{Galaxy structure}

Pursuing structural variation with mass,
Ciotti \etal (1996) considered spherical, non-rotating, two-component models
to find trends along the FP in orbital radial anisotropy, dark matter and luminous matter profiles.
Mould (2014) compared the mass profile M(r) and luminosity profile L(r) with reference to
the Jeans equation, whose three terms, 
 the density derivative,
the velocity dispersion derivative and $\beta$, relate to concentration, dispersion profile and anisotropy.

A new chapter on scaling relations is being opened by the Ultra Diffuse Galaxies. The difficulty of measuring their velocity dispersions means that few have been placed in the FP. Their formation may be environment dependent (Forbes et al 2020). Their dark matter content may be different from the standard halo mass / stellar mass relation.

\subsubsection{Concentration}
Sersic indices have been measured from 2MASS data for the Campbell \etal (2014) 6dFGS sample.
Figure 3 is the deviation from the virial plane as a function of Sersic index.
A list of 129 galaxies was excluded, usually the result of missing parts of the
image in the original 2MASS processing or close contamination
by another galaxy or a star. 
\begin{figure}
\includegraphics[scale=0.35, angle=-90]{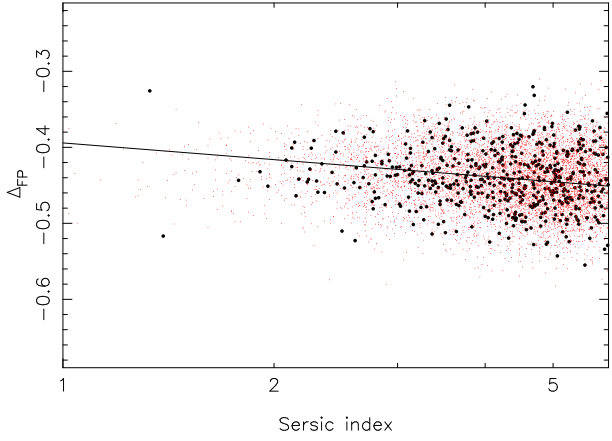}
\caption{Deviation from the virial plane. 
To separate resolution effects, the apparently largest 541 galaxies
are filled black symbols and the line is a least squares fit.}
\end{figure}

The strong dependence of tilt on stellar mass is not repeated with Sersic index.
The trend is weaker than that seen by D'Onofrio \etal in their WINGS-Sample II.

\subsubsection{Dispersion profile}
Single fibres were allocated to galaxies in 6dFGS, and central velocity dispersions
for the galaxies are all that is available. The SAMI survey, however, (Croom \etal 2012)
offers fibre bundle resolution on a 6dFGS sized sample of galaxies.
As this technology takes over from slit or single fibre spectroscopy, it will allow us 
to pursue the influence of the remaining terms in the Jeans equation on deviation from the virial plane, an exciting development.
\subsubsection{Anisotropy}

\begin{figure}
\includegraphics[scale=0.35, angle=-90]{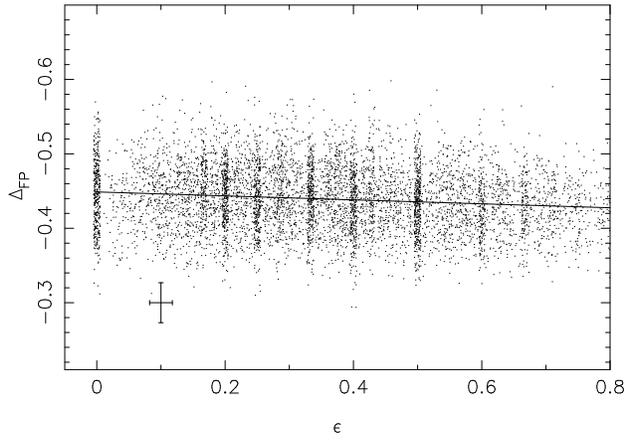}
\caption{Deviation from the virial plane vs ellipticity (1-r),
where 1/r is the ratio of the major and minor axes. The horizontal error
is the mean rounding error in these quantities, and a lower limit on the 
observational uncertainty. Popular values of ellipticity have been slightly blurred for improved visibility.}
\end{figure}

Anisotropic kinematics have been explored in many papers on triaxiality
(e.g. Binney 1987). Anisotropy manifests itself in crudely ellipticity and
one can look for dependences. D'Onofrio \etal in their WINGS-Sample II also see a trend with ellipticity.
The majority of our 6dFGS galaxies have axial ratios given by NED.
We see this trend too in Figure 4, but the slope is 0.03, rather than 0.24.

\subsection{Halo mass}
Behroozi \etal (2010) developed a relation between stellar mass and halo mass based on mass assignment
of dark matter halos in simulations to stellar masses in standard luminosity functions.
We have used their equation 21 with their standard parameters\footnote{With a 1$\sigma$ reduction in M$_{*0}$} for zero redshift to plot Figure 5. 
The tilt is reduced 
by an order of magnitude from Figure 2, indicating that the stellar mass / halo mass relation is crucial to setting the tilt of the FP.

\begin{figure}
\includegraphics[scale=0.35, angle=-90]{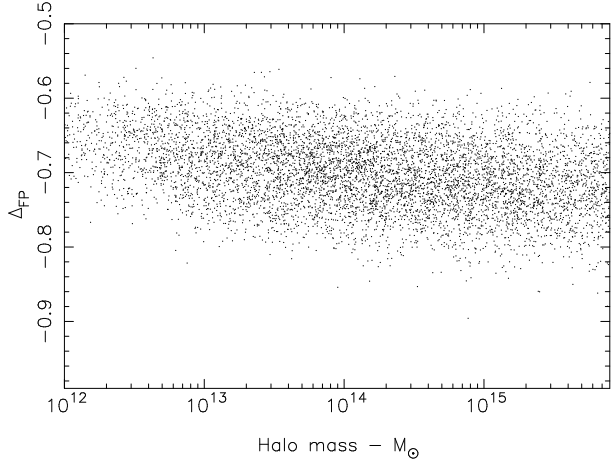}
\caption{Deviation from the virial plane versus halo mass.
The tilt is reduced by an order of magnitude from Figure 2.}
\end{figure}

\section{Star formation history and Chemical enrichment}
It is known that there are strong stellar population
trends with velocity dispersion for red sequence galaxies, e.g. Nelan \etal (2005).
Higher velocity dispersion red-sequence galaxies are, on average, older, more
metal-rich, and more alpha-enhanced than lower velocity dispersion galaxies.
Allanson \etal (2009)
examined how these trends and different star formation histories could
give the tilt of the FP. 
They found that the FP tilt and scatter are driven
primarily by stellar age effects and also concluded that
``the contribution of stellar populations to the tilt
of the fundamental plane is highly dependent on the
assumed star formation history (SFH)."
Fitzpatrick \& Graves (2014), on the other hand, find that to zeroth-order the SFH
of their early-type galaxy (ETG) sample is fully
captured by the structural parameters $\sigma$ and I$_e$, and any
differences in the SFH with environment
at fixed structure are only slight. The SFH-structure correlation they
 observe constrains the degree to which late-time evolutionary processes 
can alter the SFHs/structures of early-type galaxies in their sample.

Following the analysis of SDSS spectra by Graves \& Faber (2010), we have calculated SB residuals from the FP, $\Delta \log SB_e ~=~\log SB_e - (\log r_e - a\log \sigma)/b $,
by comparing the observed SB of each galaxy with the Magoulas et al (2012) expectation (equation 5). High SB can come about
from high stellar density or relatively recent star formation or both.
Figure 6 \& 7 show that 6dFGS spectra lead us to concur with Graves \& Faber (2010) on  that point. Graves \& Faber go on to correlate
stellar mass / halo mass ratio with population characteristics. Population
correlations in 6dFGS lead in a somewhat different direction.

The stellar populations of early-type galaxies form a two dimensional family in $\sigma$ and SB. [Fe/H] and [$\alpha$/Fe] increase\footnote{$\alpha$ refers to $\alpha$ elements, such as Mg, Ca, \& Ti. The bracket notation is the conventional element number density relative to that of the Sun.} with $\sigma$. In the other dimension age and [$\alpha$/Fe] decrease with SB, while [Fe/H] increases.
At fixed $\sigma$, galaxies with younger ages, lower [$\alpha$/Fe], and greater total metal enrichment also have larger SB and formed more stars.
Enhanced [$\alpha$/Fe] ratios  imply that enrichment proceeded by mostly core-collapse supernovae. 
and that this star formation was of short duration.

Figure 8 shows the correlation between SB excess and age at every value of $\sigma$.
SB excess can arise through star formation modifying M/L or a mass density excess.
For a Salpeter IMF fading of a billion year old burst of star formation occurs at the rate of $\Delta \log SB ~\approx~ \Delta \log t$. The vertical pillars in Figure 8 correspond to  $\Delta \log SB ~\gapp~ 2 \Delta \log t$.
Either the fading is fast due to a steeper than Salpeter IMF in the 1--2 M$_\odot$ regime or there is also present
a mass density excess in the SB excess galaxies. Following our model in $\S$2, galaxies with more than average dissipation
(small $r_e$) could achieve the observed SB excess in this way without an abnormal IMF.

It is also important to note that the age of a stellar population measured here really refers to the date
of the most recent burst of star formation.
 Graves \& Faber present a ``staged star formation" model
in which there are systematic variations in the duration of star formation at fixed $\sigma$.
At fixed $\sigma$, galaxies with longer duration star formation have higher SB,
and higher total stellar mass 
at fixed effective radius.
~They also are more enhanced in both Fe and $\alpha$-elements, with lower [$\alpha$/Fe] than their counterparts that experienced a shorter run of star formation. This could be due to higher ``conversion efficiencies" of turning gas into stars. Galaxies with low metallicity may have truncated their SFH due to feedback.


A new aspect 
is the realisation that the stellar initial mass function (IMF) probably depends on velocity dispersion, \eg
       Spiniello \etal (2013). Tortora \etal (2013) see
``a clear trend of steepening IMF with $\sigma$"
and this is also seen (apart from one anomalous galaxy) in Figure 8 of Smith \& Lucey (2013). 
This trend changes the stellar population contribution to the FP tilt, \eg
Grillo \& Gobat (2010).
Smith (2014) worries that one (or both) of the dynamical and spectroscopic constraints on the IMF has not accounted fully for the main confounding factors, element abundance ratios and dark matter contributions.~Clearly more work needs to be done on this aspect of the FP, with due recognition of degeneracy between dark matter fraction and IMF.

Dutton \& Treu (2014) and Lagattuta et al (2017) find that a `bottom heavy' Salpeter-like IMF is preferred for massive galaxies over the light Chabrier-like IMFs usually preferred for Milky Way-type galaxies\footnote{Chabrier (2003)}. 
The line indices measured by Campbell et al (2014) in 6dFGS are not IMF sensitive, and we cannot comment from this perspective on its influence on the FP ($cf.$ Mould 2014).

The FP can also be studied as a function of redshift and the evolution of early type galaxy radius
is seen in Naab et al (2009) (models) and Papovich et al (2012) (deep survey data), for example.
In due course this approach will be immensely helpful in understanding the FP, but a large
sample of z $\sim$ 1 velocity dispersions is not yet available (cf. Cerulo et al 2014).


\begin{figure}
\includegraphics[scale=0.35, angle=-90]{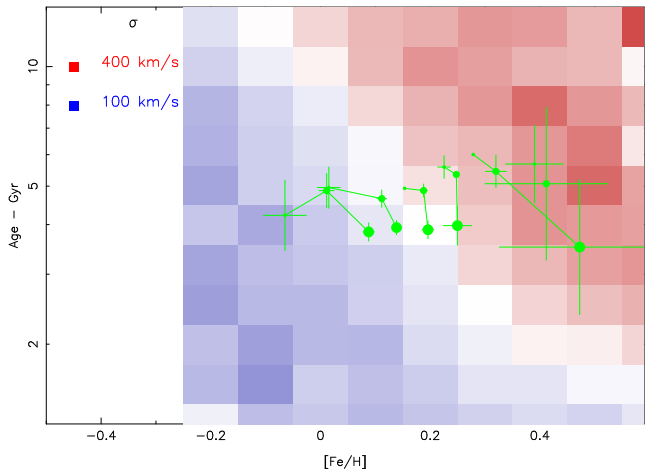}
\caption{Location of 6dFGS galaxies in the metallicity, age plane.
Properties are colour coded by velocity dispersion with giant (red) galaxies
old and metal rich and lesser (blue) galaxies younger and metal poorer.
Galaxies which are low surface brightness relative to the mean
for their $\sigma$ and r$_e$ on the FP are small green symbols;
high SB galaxies are large green symbols. Green lines connect bins 
characterized by their range in $\sigma$. An upward vertical green line would be a 
stellar population fading at constant metallicity.}
\end{figure}

\begin{figure}
\includegraphics[scale=0.45, angle=-90]{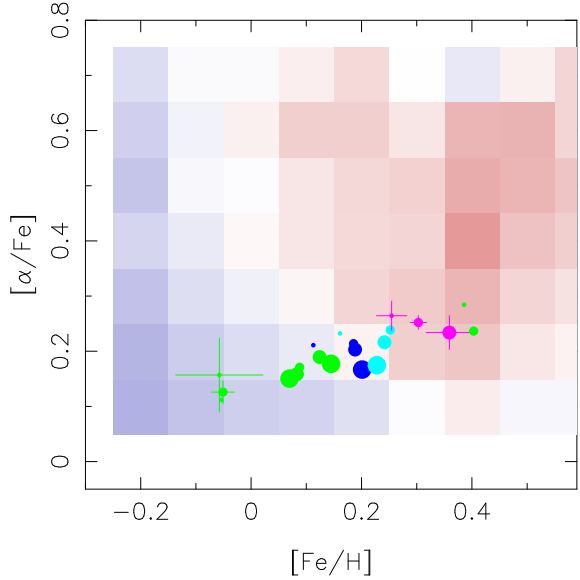}
\caption{Location of 6dFGS galaxies in the metallicity, $\alpha$-richness plane.
The giant/dwarf red/blue $\sigma$ colour coding is that of the previous figure.
Symbols are also sized by surface brightness as in the previous figure,
but the intermediate $\sigma$ bins are colour coordinated, rather than connected. 
Green $<$ 150 km/s; blue is 150--250 km/s; aqua is 250--350 km/s;
pink $>$ 350 km/s.
At a given $\sigma$ $\alpha$-richness tends to anti-correlate with metallicity. More colours are needed to distinguish the four green constant $\sigma$ bins, which show the same effect, but this is not apparent, as they are all green.}
\end{figure}

\begin{figure}
\includegraphics[scale=0.3, angle=-90]{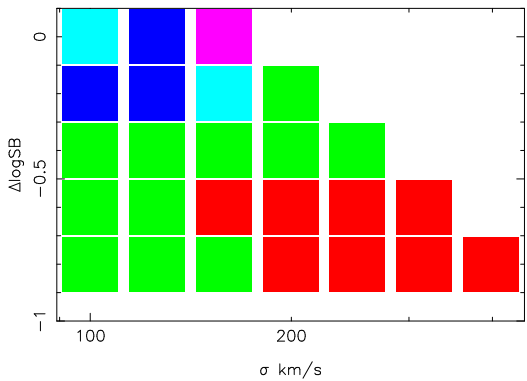}  
\caption{SB excess over the FP as a function of age and $\sigma$.
Colour increments from red through blue are each 0.11 dex in age.
Vertical pillars, blue towards the top, are a stellar population fading at constant velocity dispersion.}
\end{figure}

\section{The Tully Fisher Relation}

Studies of the relation between rotation velocity or 21 cm profile width
and disk galaxy luminosity have been carried out both in clusters of
galaxies and in the field (Tully \& Fisher 1977; Masters et al 2008).
In clusters loaded with x-ray emitting hot gas the disks can be depleted of gas,
spoiling the rotation velocity measurement (Teerikorpi et al 1992).
In the field there is a requirement
for accurate distances, as found locally in galaxies with Cepheids (Sakai et
al 2000; Mould \& Sakai 2008).

Aaronson et al (1979) pointed out that some of the complexity in a halo mass,
spiral galaxy luminosity relation can be avoided in the infrared. Aaronson
et al (1983) did not find a morphological type dependence of the infrared Tully
Fisher relation, but when bulge velocity dispersion is substituted for type,
a second parameter in the scaling relation does become apparent (Cortese et
al 2014), Tonini et al 2014).

Ultra diffuse galaxies (van Dokkum et al 2015) have now arrived at an opportune time. There are clearly different processes involved in forming them from those that make normal galaxies and shape the well populated plane. Discover those processes and we shall be adding physically to the understanding of galaxy formation, rather than just achieving a good mapping to plausible semi-analytic models of what is happening ``sub-grid''.

The dynamical understanding of UDGs is at a very early stage, however. There are perhaps a dozen, but more properly half a dozen UDGs, fully investigated kinematically (e.g. van Dokkum et al 2019, Forbes et al 2019). It is apparent that in the projection of the FP (Faber-Jackson or Tully-Fisher\footnote{Faber \& Jackson (1976), Tully \& Fisher (1977)}) UDGs are overluminous (Mancera Pina et al 2019) for their dynamical parameter ($\sigma$ or rotation) and that this can arise simply from their deviation in surface brightness, the virial theorem referenced to the surface brightness law of normal galaxies, Equation 5 of Aaronson, Huchra \& Mould (1979) 
 is $$V_{max}^4 \propto \mu_0 M$$ where the LHS (velocity) is the maximum rotation velocity of the disk, $\mu_0$ is its central surface brightness and M is the mass of the galaxy. Clearly, for a given rotation velocity, low surface brightness is compensated by high mass and elevated luminosity for constant M/L.

However, choosing between different theories of how UDGs came to be where they are in the FP will require both accurate placement of a significant sample in the FP, and development of theories such as Silk's (2019) mini-bullet cluster notion (for removal of dark matter), formation in clusters (Sales et al 2019), formation in groups (Jiang et al 2019) and tidal heating (Burkert 2017; Carleton et al 2019) to locate them in the FP.
\subsection{Eagle simulations of the TF relation}

A development that is full of promise for understanding scaling relations is
the advent of galaxy evolution simulations with baryons (Lagos et al 2016, Brooks et al 2017).
Figure 9 shows the gas distribution in a galaxy with stellar mass 10$^{10}~M_\odot$.
\begin{figure}
\includegraphics[scale=0.3, angle=-90]{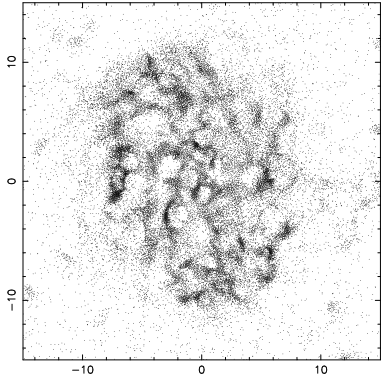}  
\caption{Neutral hydrogen particles in a disk galaxy in the Eagle simulations.
Observed HI images can be compared, such as the LMC (Kim et al 2007), which shows such structure, and M31 (Braun 1990).}
\end{figure}

To plot the rotation curve of the gas particles in Figure 10, it is necessary to find the rotation axis, and this was done by rotating the disk to edge on. This can be done by operating on the Eagle datacube phase space using rotation matrices or quaternions.
\begin{figure}
\includegraphics[scale=0.3, angle=-90]{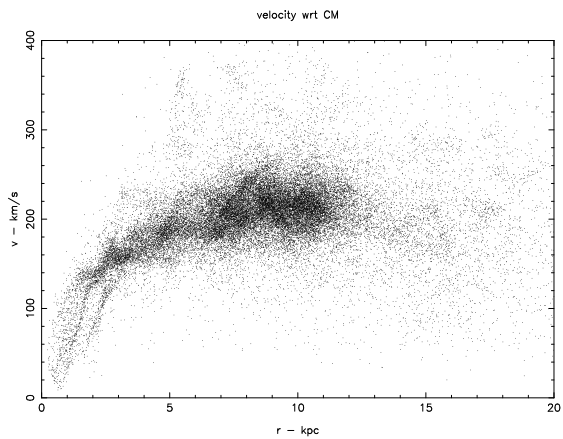}  
\caption{The rotation curve of the gas particles in the previous figure.}
\end{figure}

 The location of the observer in an Eagle datacube is arbitrary.
Therefore, we can obtain a 21 cm profile of the gas particles from infinity in any radial direction of the disk. Two views are shown in Figures 11 \& 12.
\begin{figure}
\includegraphics[scale=0.3, angle=-90]{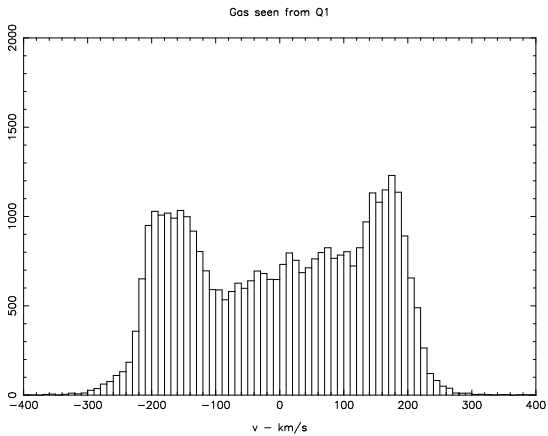}  
\caption{Neutral hydrogen profile viewed from one of the disk axes.}
\end{figure}

\begin{figure}
\includegraphics[scale=0.3, angle=-90]{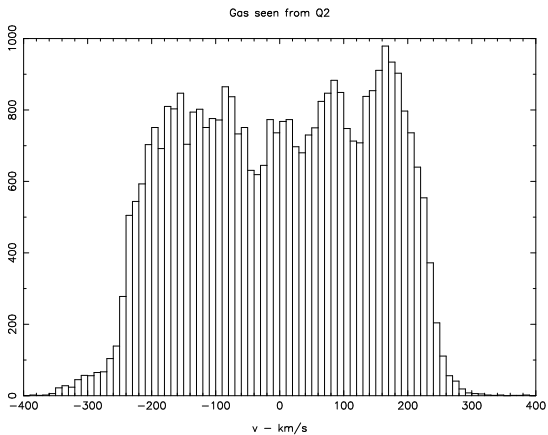}  
\caption{Neutral hydrogen profile viewed from the orthogonal axis.}
\end{figure}

A Tully-Fisher relation can be predicted from these simulations at a variety
of rotation velocities. Residuals from the Tully Fisher relation
can be studied for second parameters, such as the velocity dispersion of the bulge stars.

Lu et al (2020) have studied the evolution of the FP in the Illustris simulations (Springet et al 2018). They find that the proportionality constant in equation (5) evolves, but not the power law coefficients ($a, b$).
\section{Promising pathways}

The most rapidly advancing insights into scaling relations are
coming from simulations. As the resolution of simulations improve, and as their subgrid physics assumptions become more realistic and more fundamental, some of the clues that appear from the observations will receive support, and other will be seen to be less relevant.

A partial list of physical influences on the FP is

\begin{itemize}
\item dissipation resulting from the merger process fundamental to the formation of early type galaxies
\item feedback resulting from merger induced star formation and other processes
\item mass to light ratio systematics including the IMF, 
metallicity and age of stellar populations and the dark matter fraction.
\end{itemize}

We have shown that the primary tilt of the FP is large and dependent on the mass of the galaxy.
Its origin is a hybrid of M/L dependence on mass and structural variation with mass.
Our principal new insight is Figure 8 where
we have also shown that galaxies with more recent star formation stand out from the FP in SB.
This can be understood in terms of efficient star formation and the additional light emitted by billion year old main sequence stars. Separation of the effects of SFH
and IMF here requires further work.

If hydrodynamic simulations of the formation of ellipticals were available
right down to the star formation scale, then theory could make 
a statement not only about dissipation and feedback but also about the $0.5 M_\odot/0.1M_\odot$ stellar mass ratio expectations
for massive and intermediate mass ellipticals. However, at present that may be a bridge too far,
and IMF issues should be pursued observationally in the near term.~


Dissipation, feedback and chemical enrichment are active development areas of the current generation of galaxy formation simulations.
Comparison of their predictions with large integral field surveys of early type galaxies, such as ATLAS$^{3D}$ (Cappellari \etal 2013ab) and SAMI,
which measure the density profile and dispersion profile terms in the Jeans equation, promise further progress
in understanding the FP, especially as regards the relationship we see in 6dFGS between structure and FP tilt.

 With the advent of new surveys DESI, Taipan (Da Cunha et al 2017), 
Wallaby (Koribalski et al 2020) and Dragonfly (Abraham 2020) we can expect a deeper and more general understanding of the dynamical equilibrium of galaxies to emerge,
and also of galaxy scaling relations that stem from them.
Combining this with galaxy simulations of the Eagle type, an exciting new era of understanding galaxy scaling relations is beginning. 

\section*{References}
Aaronson, M., Huchra, J., \& Mould, J. 1979, {\it The infrared luminosity/H I velocity-width relation and its application to the distance scale.} ApJ, 229,1-13 \\
Aaronson, M.,  \& Mould, J. 1983, {\it A distance scale from the infrared magnitude/H I velocity-width relation. IV. The morphological type dependence and scatter in the relation; the distances to nearby groups.} ApJ, 265,1-17 \\
Abraham, R. 2020, {\it The Dragonfly Telephoto Array: how it works and where it is going} AAS 2354401\\
Allanson, S., Hudson, M., Smith, R. J. \& Lucey, J. R. 2009, {\it The Star Formation Histories of Red-Sequence Galaxies, Mass-to-Light Ratios and the Fundamental Plane} ApJ, 702, 1275-1296	\\
Behroozi, P. S.,  Conroy, C. \& Wechsler, R. 2010,{\it A Comprehensive Analysis of Uncertainties Affecting the Stellar Mass-Halo Mass Relation for 0 $<$ z $<$ 4} ApJ, 717, 379-403\\
Binney, J. 1987, {\it Observable Consequences of Triaxial Halos} IAUS, 117, 303\\
Brooks, A., Papastergis, E., Christensen, C.R., Governato, F. Stilp, A.,
 Quinn, T. R. \etal 2017, {\it How to Reconcile the Observed Velocity Function of Galaxies with Theory} ApJ, 850, 97\\
Busarello, G. 1997, {\it The relation between the virial theorem and the fundamental plane of elliptical galaxies.}, A\&A, 320, 415-420\\
Braun, R. 1990, {\it The Interstellar Medium of M31. I. A Survey of Neutral Hydrogen Emission} ApJS, 72, 755\\
Burkert, A. 2017, {\it The Geometry and Origin of Ultra-diffuse Ghost Galaxies} ApJ, 838, 93\\
Carleton, T., Errani, R., Cooper, M., Kaplinghat, M., Penarrubia, J.,
 Guo, Yicheng et al 2019, {\it The formation of ultra-diffuse galaxies in cored dark matter haloes through tidal stripping and heating} MNRAS, 485, 382-395\\
Campbell, L., Lucey, J., Colless, M., Jones, D.H., Springob, C., Magoulas, C. \etal 2014, {\it The 6dF galaxy survey: fundamental plane data} MNRAS, 443, 1231\\
Cappellari, M., Scott, N., Alatalo, K., Blitz, L., Bois, M. Bournaud, F,
\etal 2013a, {\it The ATLAS$^{3D}$ project - XV. Benchmark for early-type galaxies scaling relations from 260 dynamical models: mass-to-light ratio, dark matter, Fundamental Plane and Mass Plane} MNRAS, 432, 1709-1741\\
Cappellari, M., McDermid, R.M. Alatalo, K., Blitz, L. Bois, M. Bournaud, F. \etal 2013b,{\it  The ATLAS<SUP$^{3D}$ project - XX. Mass-size and mass-$\sigma$ distributions of early-type galaxies: bulge fraction drives kinematics, mass-to-light ratio, molecular gas fraction and stellar initial mass function}	MNRAS, 432, 1862-1893\\
Cerulo, P.,  Couch, W. J., Lidman, C., Delaye, L., Demarco, R.,
 Huertas-Company, M. et al 2014, {\it The morphological transformation of red sequence galaxies in the distant cluster XMMU J1229+0151} MNRAS, 439, 2790-2812\\
Chabrier, G. 2003, {\it Galactic Stellar and Substellar Initial Mass Function} PASP, 115, 763-795\\
Ciotti, L., Lanzoni, B. \& Renzini, A. 1996, {\it The tilt of the fundamental plane of elliptical galaxies - I. Exploring dynamical and structural effects} MNRAS, 282, 1-12\\
Conroy, C. \& van Dokkum, P. 2012, {\it The Stellar IMF in ETGs from absorption line spectroscopy II Results}, ApJ, 760, 16\\
Conroy, C., Dutton, A., Graves, G. et al 2013, {\it Dynamical vs stellar masses in compact ETGs: Further evidence for systematic variation in the stellar IMF}, ApJ, 776, L26\\
Cortese, L., Fogarty, L. M. R., Ho, I. -T., Bekki, K., Bland-Hawthorn, J.,
 Colless, M. et al 2014, {\it The SAMI Galaxy Survey: Toward a Unified Dynamical Scaling Relation for Galaxies of All Types} ApJL, 795, L37\\ 
Croom, S. M.,  Dawe, J., Fiegert, K., Frankcombe, L. \etal 2012,{\it The Sydney-AAO Multi-object Integral field spectrograph} MNRAS, 421, 872-893   \\
da Cunha, E.,  Hopkins, A., Colless, M., Taylor, E. N., Blake, C.,
 Howlett, C. et al 2017, {\it The Taipan Galaxy Survey: Scientific Goals and Observing Strategy} PASA, 34, 47\\
D'Onofrio, M.,  Fasano, G., Moretti, A., Marziani, P., Bindoni, D.,
 Fritz, J. \etal 2013, {\it The hybrid solution for the Fundamental Plane} MNRAS, 435, 45-63\\
Dutton, A. \& Treu, T. 2014, {\it The bulge-halo conspiracy in massive elliptical galaxies: implications for the stellar initial mass function and halo response to baryonic processes} MNRAS, 438, 3594-3602	\\
Faber, S. \& Jackson, R. 1976, {\it Velocity dispersions and mass-to-light ratios for elliptical galaxies.} ApJ, 204, 668-683 \\
Ferrarese, L. \& Merritt, D. 2000, {\it A Fundamental Relation between Supermassive Black Holes and Their Host Galaxies} ApJ, 539, L9-L12\\	
Fitzpatrick, P. \& Graves, G. 2014, {\it Early-Type Galaxy Star Formation Histories in Different Environments} astro-ph 1403.7836\\
Forbes, D., Alabi, A., Romanowsky, A.J., Brodie, J., Arimoto, N.
et al 2019, {\it An ultra diffuse galaxy in the NGC 5846 group from the VEGAS survey} A\&A, 626, 66\\
Forbes, D., Martin, C., Matuszewski, M.,  Romanowsky, A.J., Villaume, A.
{\it The formation of ultradiffuse galaxies in clusters} et al 2020, MNRAS, 492, 4874-4883\\
Graham, A. 2012, {\it Extending the $M_{BH},\sigma$ diagram with dense nuclear clusters}MNRAS, 422, 1586\\	
Graves, G. \& Faber, S. 2010, {\it Dissecting the Red sequence. III. Mass-to-Light Variations in Three-dimensional Fundamental Plane Space} ApJ, 717, 803-824\\
Grillo, C. \& Gobat, R.	2010, {\it On the initial mass function and tilt of the fundamental plane of massive early-type galaxies} MNRAS, 402, L67-L71\\
Jiang, F., Dekel, A., Freundlich, J. Romanowsky, A. et al 2019, {\it Formation of ultra-diffuse galaxies in the field and in galaxy groups} MNRAS, 487, 5272-5290\\
Jones, D. H., Read, M., Saunders, W., Colless, M., Jarrett, T.
 Parker, Q. et al 2009, {\it The 6dF Galaxy Survey: final redshift release (DR3) and southern large-scale structures} MNRAS, 399, 683-698\\
Kim, S. \& Park C. 2007, ApJ, 663, 244\\
Koribalski, B., Staveley-Smith, L., Westmeier, T., Serra, P., Spekkens, K.,
 Wong, O. I. et al 2020, {\it WALLABY -- An SKA Pathfinder HI Survey}arxiv 200207311\
Kormendy, J. \& Ho, L. 2013, {\it Coevolution (Or Not) of Supermassive Black Holes and Host Galaxies} ARAA, 51, 511-653\\
Kormendy, J. \& Kennicutt, R. 2004, {\it Secular Evolution and the Formation of Pseudobulges in Disk Galaxies} ARAA, 42, 603-683\\
Lagattuta, D., Mould, J., Forbes, D., Monson, Pastorello, \& Persson, S.E. 2017, {\it Evidence of a Bottom-heavy Initial Mass Function in Massive Early-type Galaxies from Near-infrared Metal Lines} ApJ, 846, 166\\
Lagos, C., Obreschkow, D., Ryan-Weber, E., Zwaan, M., Kilborn, V., Bekiaris, G.
et al 2016, {\it The Fundamental Plane of star formation in galaxies revealed by the EAGLE hydrodynamical simulations} MNRAS, 459, 2632-2650\\
Lu, Shengdong,  Xu, Dandan, Wang, Yunchong, Mao, Shude, Ge, Junqiang et al 2020, MNRAS, 492, 5930-5939\\
Magorrian, J., Tremaine, S., Richstone, D., Bender, R., Bower, G.,
 Dressler, A. \etal 1998, {\it The Demography of Massive Dark Objects in Galaxy Centers} AJ 115, 2285-2305\\
Magoulas, Lucey, J., Lagos, C., Kuehn, K., G. Barat, D. Bian, Fuyan \etal 2012, {\it The 6dF galaxy survey: the near infrared fundamental plane of early type galaxies} MNRAS, 427, 245\\
Mancera Pina, Fraternali, F., Adams, E,, Marasco, A.,  Oosterloo, T.,
 Oman, K. et al 2019, {\it Off the Baryonic Tully-Fisher Relation: A Population of Baryon-dominated Ultra-diffuse Galaxies} ApJ, 883, L33\\
Maraston, C. 2005,{\it Evolutionary population synthesis: models, analysis of the ingredients and application to high-z galaxies} MNRAS, 362, 799-825\\
Masci, F. Cutri, R., Francis, P.,  Nelson, B. Huchra, J. \etal 2010, {\it The Southern 2MASS Active Galactic Nuclei Survey: Spectroscopic Follow-up with Six Degree Field} PASA, 27, 302-320\\
Masters, K., Springob, C. \& Huchra, J. 2008 {\it 2MTF. I. The Tully-Fisher Relation in the Two Micron All Sky Survey J, H, and K Bands}  AJ, 135, 1738-1748\\	
Moster, B., Somerville, R., Maulbetsch, C. van den Bosch, F., Maccio, A., Thorsten, N. et al 2010,{\it Constraints on the relationship between stellar mass and halo mass at low and high redshift} ApJ, 710, 903\\
Mould, J. 2014, $in$ CoEPP Tropical Workshop 2013,\\ 
{\it What are we missing in elliptical galaxies ?} arxiv 1403.1623\\
Mould, J. 2017, {\it Modified gravity and large scale flows, a review} ApSS, 362, 25\\
Mould, J. \& Sakai, S. 2008, {\it The Extragalactic Distance Scale without Cepheids} ApJL, 686, L75\\
Naab, T., Oosterloo, T., Sarzi, M., Serra, P., Weijmans, A., Young, L. 2009,{\it Minor Mergers and the Size Evolution of Elliptical Galaxies} ApJ, 699, L178-L182\\
Nelan, J.Smith, R., Hudson, M., Wegner, G. \etal 2005, {\it NOAO Fundamental Plane Survey. II. Age and Metallicity along the Red Sequence from Line-Strength Data} ApJ, 632, 137-156\\
Papovich, C., Bassett, R., Lotz, J., van der Wel, A., Tran, K. -V., Finkelstein, S. et al 2012,{\it CANDELS Observations of the Structural Properties of Cluster Galaxies at z = 1.62} ApJ, 750, 93\\
Proctor, R., Colless, M., Jones, D. H., Kobayashi, C., Campbell, L.,
 Lucey, J. \etal 2008, {\it The effects of stellar populations on galaxy scaling relations in the 6dF Galaxy Survey} MNRAS, 386, 1781-1796 \\
Sakai, S., Mould, J., Hughes, S., Huchra, J., Macri, L., Kennicutt, R. et al 2000,{\it The Hubble Space Telescope Key Project on the Extragalactic Distance Scale. XXIV. The Calibration of Tully-Fisher Relations and the Value of the Hubble Constant} ApJ, 529, 698-722\\
Sales, L., Navarro, J., Pena{\~n}afiel, L., Peng, E., Lim, S. \& Hernquist, L. 2019, {\it  
The formation of ultradiffuse galaxies in clusters}, MNRAS, 494, 1848-1858\\
Silk, J. 2019, {\it Ultra-diffuse galaxies without dark matter} MNRAS, 488, L24-L28\\
Skrutskie, M., Cutri, R., Stiening, R., Weinberg, M., Schneider, S.,
 Carpenter, J. \etal 2006, {\it The Two Micron All Sky Survey (2MASS)} AJ, 131, 1163-1183\\
Smith, R. \& Lucey, J. 2013, {\it A giant elliptical galaxy with a lightweight initial mass function} MNRAS, 434, 1964-1977\\
Smith, R. 2014, {\it Variations in the initial mass function in early-type galaxies: a critical comparison between dynamical and spectroscopic results.} MNRAS, 443, L69-L73\\
Spiniello, C.,  Trager, S., Koopmans, L., Conroy, C, \etal 2014 {\it The stellar IMF in early-type galaxies from a non-degenerate set of optical line indices} MNRAS, 438, 1483-1499 \\
Springel, V. Wang, Y. Vogelsberger, M.,  Naiman, J., Hernquist,  Sales. L.
et al 2018, {\it Redshift evolution of the Fundamental Plane relation in the IllustrisTNG simulation} MNRAS, 475, 676-698\\
Springob, C, Magoulas, C, Mould, J. \etal 2012, {\it The 6dF Galaxy Survey: stellar population trends across and through the Fundamental Plane} MNRAS, 420, 2773-2784\\
Teerikorpi, P., Bottinelli, L., Gouguenheim, L.\& Paturel, G. 1992, {\it Investigations ofthe Local Supercluster velocity field. I. Observations close to Virgo, using Tully-Fisher distances and the Tolman-Bondi expanding sphere.} A\&A, 260, 17-32\\
Tonini, C., Jones, D. H., Mould, J., Webster, R. Danilovich, T. \&
 Ozbilgen, S. 2014, {\it The fundamental manifold of spiral galaxies: ordered versus random motions and the morphology dependence of the Tully-Fisher relation} MNRAS, 438, 3332-3339\\
Tortora, C., La Barbera, F., Napolitano, N., de Carvalho, R. R.,
 Romanowsky, A. \etal 	2012, {\it An Inventory of the Stellar Initial Mass Function in Early-type Galaxies} ApJ, 765, 8\\
Tortora, C. \etal 2013, {\it SPIDER - VI. The central dark matter content of luminous early-type galaxies: Benchmark correlations with mass, structural parameters and environment} MNRAS, 425, 577-594\\
Tully, R.B. \& Fisher, R. 1977, {\it A new method of determining distance to galaxies.} A\&A, 54, 661\\
van Dokkum, P., Romanowsky, A., Abraham, R. Brodie, J. Conroy, C., Geha, M. 
et al 2015, {\it Spectroscopic Confirmation of the Existence of Large, Diffuse Galaxies in the Coma Cluster} ApJ, 804, 26\\
van Dokkum, P., Wasserman, A., Danieli, S., Abraham, R. Brodie, J.,
 et al 2019, 
{\it Spatially Resolved Stellar Kinematics of the Ultra-diffuse Galaxy Dragonfly 44. I. Observations, Kinematics, and Cold Dark Matter Halo Fits} ApJ, 880, 91\\
Walker, M., Mateo, M., Olszewski, E., Penarrubia, J., Evans, N., Gilmore, G. et al 2009, {\it A Universal Mass Profile for Dwarf Spheroidal Galaxies?}ApJ, 704, 1274-1287\\
Wolf, J. 2010, {\it Modeling mass independent of anisotropy: A comparison between Andromeda and Milky Way satellites} Highlights in Astronomy, 15, 79\\







\section*{Funding}
I am grateful for Australian Research Council (ARC) grants DP1092666 \& LP130100286 in partial support of this work.
Our dark matter research is supported by the ARC through CDMPP{\footnote{www.darkmatter.org.au}. 

\section*{Acknowledgments}
I thank MIAPP, the Munich Institute for Astro and Particle Physics
for inviting me to their workshop ``Galaxy Evolution in a New Era of HI Surveys''. MIAPP is funded by the DFG under Germany's Excellence Strategy -- EXC-2094-390783311. I am grateful to Edoardo Tescari for assistance with Eagle datacubes, and anonymous referees for recommending improvements.

\section*{Data Availability Statement}
Data from 6dFGS are archived at the Widefield Astronomy Unit of the Royal Observatory Edinburgh and Data Central at Macquarie University. This study has also used 2MASS data provided by NASA-JPL.

\end{document}